\documentclass[5p]{elsarticle}

\usepackage{lineno,hyperref}

\usepackage{amsmath}
\hypersetup{colorlinks,allcolors=black}
\journal{International Journal of Mass Spectrometry}

\begin{document}

\begin{frontmatter}

\title{High-precision mass measurement of \texorpdfstring{$^{168}$}{Lg}Yb for verification of nonlinear isotope shift}

\author[a]{D.A.~Nesterenko}
\author[a]{R.P.~de~Groote}
\author[a]{T.~Eronen}
\author[a]{Z.~Ge}
\author[a,b]{M.~Hukkanen}
\author[a]{A.~Jokinen}
\author[a]{A.~Kankainen}

\address[a]{University of Jyv\"askyl\"a, P.O. Box 35, FI-40014 University of Jyv\"askyl\"a, Finland}
\address[b]{Centre d'\'Etudes Nucl\'eaires de Bordeaux-Gradignan, 19 Chemin du Solarium, CS 10120, F-33175 Gradignan, France}

\begin{abstract}
The absolute mass value of $^{168}$Yb has been directly determined with the JYFLTRAP Penning trap mass spectrometer at the Ion Guide Isotope Separator On-Line (IGISOL) facility. A more precise value of the mass of $^{168}$Yb is needed to extract possible signatures of beyond standard model physics from high-precision isotope shift measurements of Yb atomic transition frequencies. The measured mass-excess value, ME($^{168}$Yb) = $-$61579.846(94) keV, is 12 times more precise and deviates from the Atomic Mass Evaluation 2016 value by 1.7$\sigma$. The impact on precision isotope shift studies of the stable Yb isotopes is discussed.
\end{abstract}

\begin{keyword}
\texttt Penning trap \sep High-precision mass spectrometry \sep Isotope shift
\end{keyword}

\end{frontmatter}


\section{Introduction}

The transition frequency of the same atomic line shifts between different isotopes of the same element due to changes in nuclear mass and size. This effect is known as the atomic isotope shift, and plays an important role in atomic and nuclear physics \cite{campbell2016laser}. Recently, high-precision isotope shift measurements were suggested to provide a means for searching for physics beyond the standard model (SM) \cite{Delaunay2017,Berengut2018}. Through a King-plot analysis, which compares the isotope shifts of different optical transitions for a series of isotopes, such new physics would manifest itself as a deviation from linearity. The first experimental reports of such high-precision King-plot tests are now becoming available \cite{Miyake2019,Counts2020}. Recently, precision measurements of the isotope shifts of the even-even stable $^{168-176}$Yb$^+$ isotopes showed a deviation from linearity at the 3$\sigma$ level \cite{Counts2020}. At present, it is not clear whether this deviation is due to SM or beyond the SM effects, so more experimental and theoretical work is ongoing.

The isotope shift between two isotopes of an optical transition is defined as $\nu_{ji} = \nu_j - \nu_i$, the difference in the transition frequencies of the two isotopes. A King plot is constructed by plotting the modified isotope shifts of two different optical transitions $\alpha$ and $\beta$ against each other. The modified isotope shift is obtained by multiplying $\nu_{\alpha ji}$ and $\nu_{\beta ji}$  with the reduced mass factor defined as $\mu_{ji} = 1/m_j - 1/m_{i}$. The atomic masses are input parameters for such a King plot, and precise mass values are hence needed. While the atomic masses of $^{170, 172, 174, 176}$Yb are very well known in the most recent Atomic Mass Evaluation 2016 (AME2016) ($\delta m = 10 - 15$~eV/$c^2$ \cite{Wang2017}), based on the measurements at the FSU Penning trap \cite{FSU2012}, the mass of $^{168}$Yb has a much higher uncertainty of 1.2~keV/$c^2$ \cite{Wang2017}. 
For the search of physics beyond the standard model via high-precision King-plot tests,  it is therefore critical to improve the precision of the $^{168}$Yb mass value.
Here we report on the most precise absolute mass value of $^{168}$Yb to date, achieved via high-precision cyclotron frequency-ratio measurements with the double Penning trap JYFLTRAP.

\section{Experimental method and results}

The JYFLTRAP double Penning trap mass spectrometer \cite{Eronen2012} is placed at the Ion Guide Isotope Separator On-Line (IGISOL) facility \cite{Moore2013}. The Yb  ions were produced in the electric discharge ion source placed inside the IGISOL target chamber. The singly charged ions were accelerated to 30 keV and then mass-separated from the natural isotopic mixture of Yb$^+$ ions using a 55$^{\circ}$ dipole magnet.
Then, the selected Yb$^+$ ions were stopped and cooled in a gas-filled radio-frequency quadrupole \cite{Nieminen2001}, which transformed the continuous beam into ion bunches. 

The ion bunches were transported to the JYFLTRAP double Penning trap. They were cooled, centered and additionally purified via a mass-selective buffer-gas cooling technique \cite{Savard1991} in the first (preparation) trap. In the second (measurement) trap the cyclotron frequency
\begin{equation} \label{eq:qbm}
\nu_{c} = \frac{1}{2\pi}\frac{q}{m}B
\end{equation}
of the ion with mass $m$ and charge $q$ in the magnetic field $B$ was measured with the phase-imaging ion-cyclotron-resonance (PI-ICR) technique \cite{Eliseev2013}. The cyclotron frequency of the ion was determined as a sum of its two radial-motion frequencies in the trap: magnetron frequency $\nu_{-}$ and modified cyclotron frequency $\nu_{+}$. The measurement scheme used for the cyclotron frequency determination has been described in detail in Ref. \cite{Eliseev2014} and its implementation at JYFLTRAP has been demonstrated in Ref. \cite{Nesterenko2018}. The PI-ICR method has already been successfully employed for high-precision $Q$-value measurements in offline \cite{Nesterenko2019} and online \cite{Roubin2020} experiments at JYFLTRAP.

The ions accumulated the ``magnetron'' and ``cyclotron'' phases of ion radial motion during the phase accumulation time $t_{acc}$ of free rotation in the trap. The ion radial motion was projected onto a position-sensitive detector (MCP detector with a delay line anode), placed outside the magnet which houses the Penning traps. The positions of the ``magnetron'' and ``cyclotron'' phase images on the detector are defined by the polar angles $\alpha_-$ and $\alpha_+$, respectively, with respect to the trap center (projection of ions from the center of the trap). The cyclotron frequency is determined as
\begin{equation} \label{eq:alpha}
\nu_{c}= \nu_{-} + \nu_{+} = (\alpha_c + 2 \pi n)/ 2 \pi t_{acc},
\end{equation}
where $\alpha_c = \alpha_+ - \alpha_-$ is the angle between two phase images, $n$ is the full number of revolutions, which the studied ions would perform in a pure magnetic field $B$ during a phase accumulation time $t_{acc}$.

The phase accumulation time used in the PI-ICR measurements was 600 ms. The phase spots and center spot were alternately accumulated during a single 5-minutes cyclotron frequency measurement. To determine the cyclotron frequency ratio $R = \nu_{c}^{ref} / \nu_{c}^{int} $, where $\nu_{c}^{ref}$ and $\nu_{c}^{int}$ are the cyclotron frequencies of the reference ion and the ion of interest, respectively, the measurements of $\nu_{c}^{ref}$ and $\nu_{c}^{int}$ were performed alternately. The frequency $\nu_{c}^{ref}$, measured before and after the measurement of $\nu_{c}^{int}$, was linearly interpolated to the time of $\nu_{c}^{int}$ measurement and a single frequency ratio $R_i$ was determined. After 3$-$5 hours of such measurements a weighted mean ratio $R_{4h}$ was calculated and the maximum of internal and external error \cite{Birge1932} was chosen. The final ratio $\overline{R}$ was calculated as a weighted mean ratio of $R_{4h}$ with the maximum of internal and external error and the atomic mass was determined as:
\begin{equation} \label{eq:mass}
m = (m_{ref} - m_{e}) \times \overline{R} + m_{e},
\end{equation}
where $m_{ref}$ is an atomic mass of reference and $m_{e}$ is an electron mass. The binding energy of the valence electron in ytterbium is several eV \cite{Lotz1970} and the measured frequency ratios $\overline{R}$ are close to unity ($\overline{R} - 1 \sim 0.01$). Thus, the contribution of the electron binding energy to the atomic mass $m$ is less than 1 eV and can be neglected.

\begin{figure}[t!]
\includegraphics[width=0.49\textwidth]{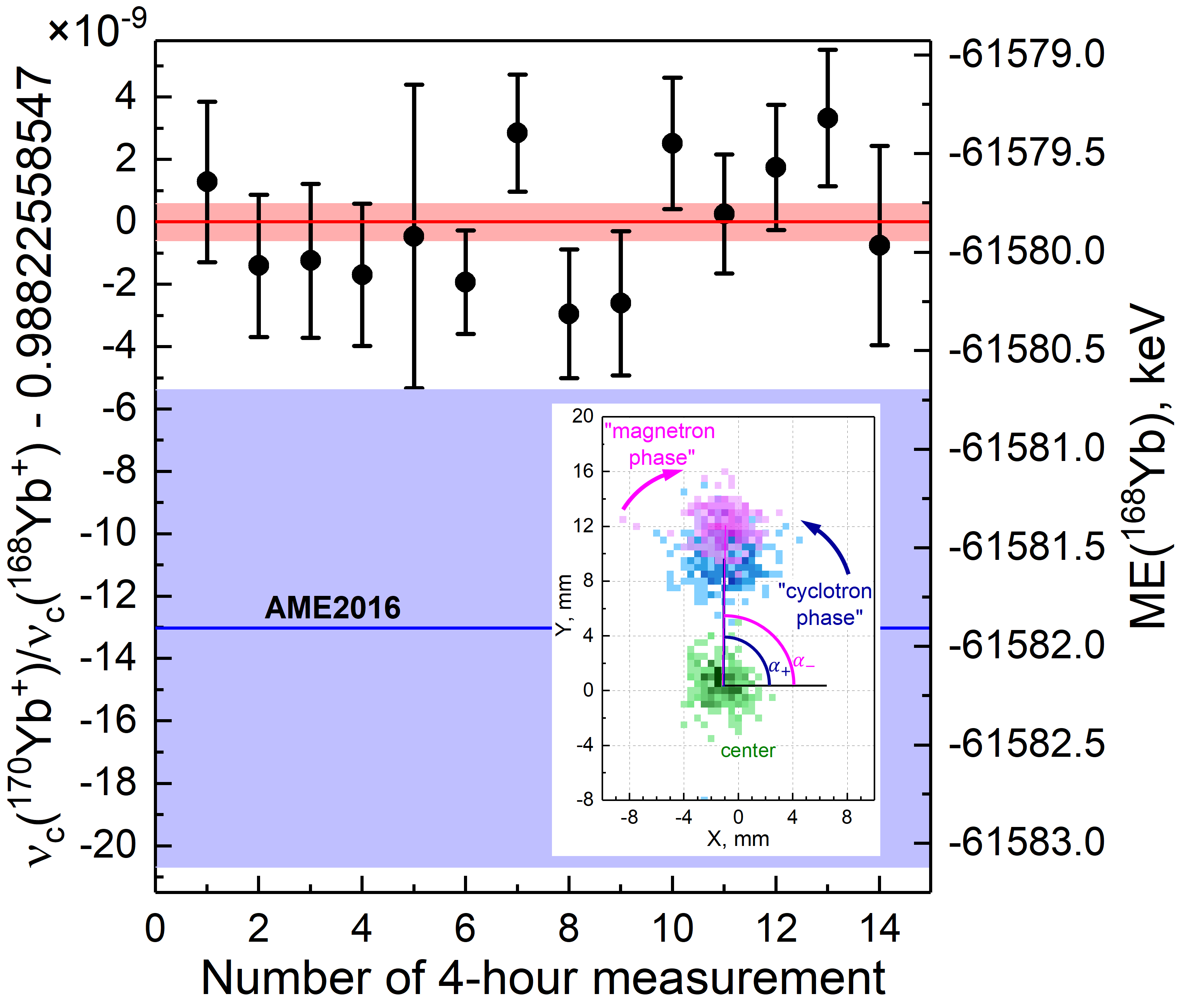}
\caption{The cyclotron frequency ratios $R_{4h} = \nu_{c}(^{170}$Yb$^+) / \nu_{c}(^{168}$Yb$^+) $ (left axis) and mass-excess values for $^{168}$Yb (right axis) determined in around 4-hour measurements. The red line and the red-shaded band represent the weighted mean value and total final uncertainty, respectively. The blue line with blue-shaded band represent the AME2016 value \cite{Wang2017} with the uncertainty. The insert shows the center and phase accumulated spots for $^{168}$Yb$^+$ on the position-sensitive detector for a single cyclotron frequency measurement (5 min) with the PI-ICR method.}
\label{fig:ratios}
\end{figure}

The uncertainty due to nonlinear drift of the magnetic field between two neighboring frequency measurements was at the level $\delta_B = 2\times10^{-12}$ min$^{-1}$ \cite{JYFLTRAP-systematics} and made a non-significant contribution to the total uncertainty. To check the possible frequency shifts due to ion-ion interactions, a count-rate class analysis \cite{Roux2013} was performed for the determined cyclotron frequency ratios. No dependence of the frequency ratio on the number of detected ions was observed, and only data with detected $1-5$ ions/bunch were taken into account in the analysis. To minimize the systematic uncertainties due to distortion of the ion-motion projection onto the detector, the positions of the magnetron and cyclotron phase spots were chosen such that the angle between them did not exceed a few degrees (see Fig.~\ref{fig:ratios}).

Altogether 14 around 4-hour measurements were performed for the cyclotron frequency ratio $R_{4h}$ =\\ $\nu_{c}(^{170}$Yb$^+) / \nu_{c}(^{168}$Yb$^+) $ (Fig.~\ref{fig:ratios}). The final weighted mean ratio with the total uncertainty is $\overline{R}$ = 0.98822558547(59) yielding a mass-excess value ME($^{168}$Yb) = $-$61579.846(94) keV.

To study possible systematic uncertainties the cyclotron frequency ratio of $^{172}$Yb$^+$ to $^{170}$Yb$^+$ was performed in a similar manner. 
The mass values for the $^{172}$Yb$-^{170}$Yb pair are known with a precision well below $10^{-10}$ \cite{Wang2017,FSU2012} and differ by two mass units as for the $^{170}$Yb$-^{168}$Yb pair, providing a cross-check for the accuracy of our measurement.
The measured cyclotron frequency ratio is $\overline{R}$ =\\ $\nu_{c}(^{172}$Yb$^+) / \nu_{c}(^{170}$Yb$^+)$ = 0.98835833483(49) yielding the mass-excess value ME($^{170}$Yb) = $-$60763.961(79) keV. Our result is in good agreement with AME2016 \cite{Wang2017} differing by $-2.6(50) \times 10^{-10}$ for the frequency ratio and $-42(80)$ eV for the mass-excess value. Therefore, no additional systematic uncertainty was added to the measured frequency ratio $\nu_{c}(^{170}$Yb$^+) / \nu_{c}(^{168}$Yb$^+) $ .

\begin{figure}[h!]
\includegraphics[width=0.49\textwidth]{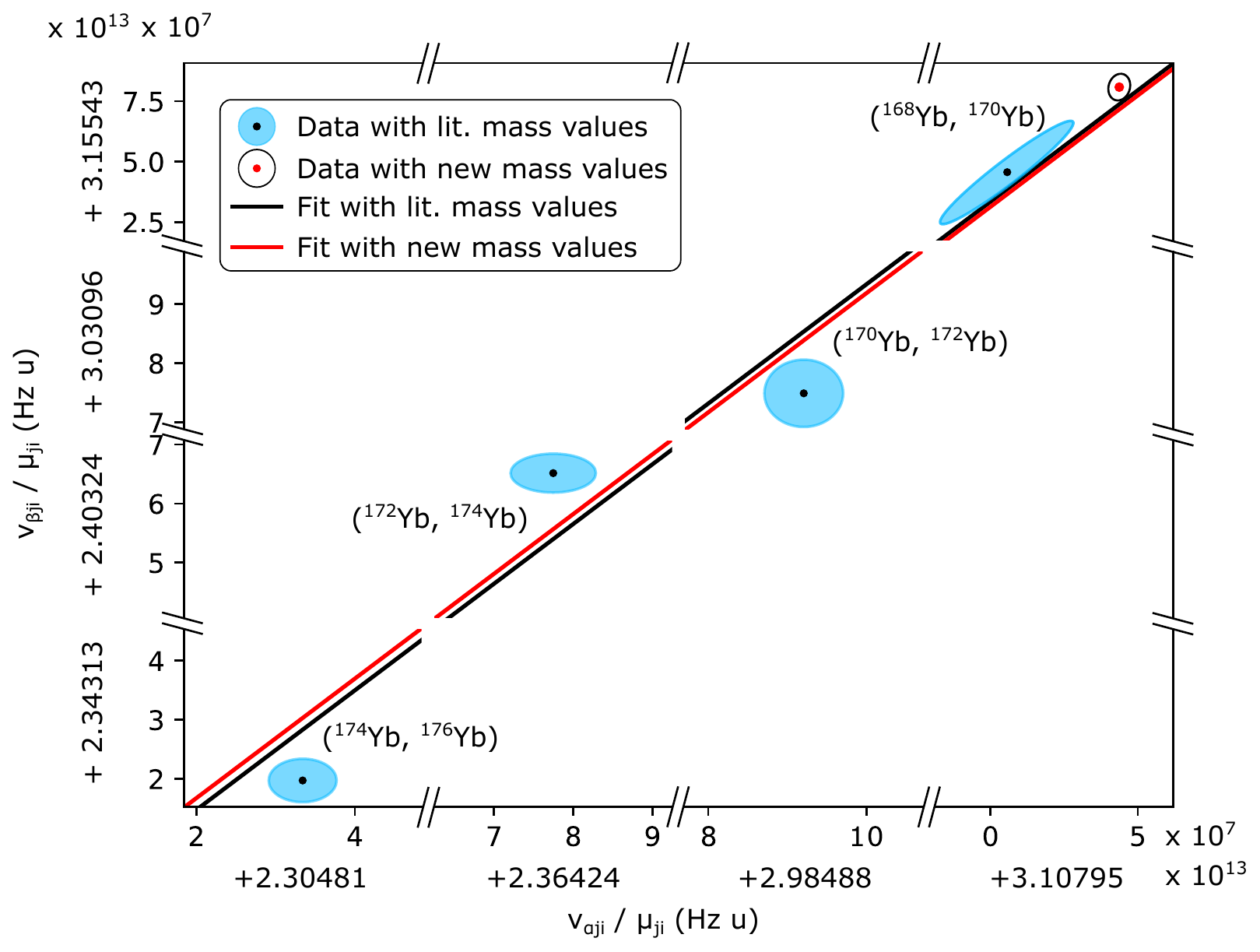}
\caption{King plot of the $\alpha=411$\,nm and $\beta=436$\,nm transitions reported in \cite{Counts2020}. Black data points and their accompanying blue shaded uncertainty ellipses are obtained using the literature mass values, while the red data point and white uncertainty ellipse is obtained with our new mass value of $^{168}$Yb. A linear fit of these points is also shown, in black for the set with literature data, and in red with the new mass value. There is no significant change in the non-linearity, though a small change in the straight line fit parameters is observed.}
\label{fig:kingplot}
\end{figure}

\section{Discussion}
The mass-excess value for $^{168}$Yb measured in this work ($-$61579.846(94) keV) differs from the AME2016 value ($-$61581.9(12) keV \cite{Wang2017}) by 2.1(12) keV, i. e., by 1.7 standard deviations, and it is 12 times more precise. The AME2016 mass value of $^{168}$Yb is based on the $Q_{\epsilon\epsilon}$-value measurement ($^{168}$Yb$-^{168}$Er) and mass of $^{168}$Er \cite{Wang2017}. The $Q_{\epsilon\epsilon}$-value was determined by Penning-trap mass-ratio measurements at SHIPTRAP \cite{Eliseev2011} with an accuracy of 250 eV. The mass value of $^{168}$Er was derived from nuclear reactions (mainly from $^{167}$Er($n$,$\gamma$)$^{168}$Er), and represents the main contribution to the mass uncertainty of $^{168}$Yb. It has already been shown that masses derived in indirect methods ($\beta-$decay spectroscopy, nuclear reactions) might be inaccurate in a broad range of mass numbers and can exhibit discrepancies with direct mass measurements \cite{Eliseev2010,Nesterenko2019}.

Using the new mass-excess value of $^{168}$Yb, we repeat the King-plot analysis presented in \cite{Counts2020}. Figure \ref{fig:kingplot} shows the King plot with both the literature mass values from AME2016 \cite{Wang2017}, and our updated mass value. The tilt of uncertainty ellipse for the $^{168}$Yb$-^{170}$Yb pair using the literature mass values is due to the correlation in the modified isotope shifts $\nu_{\alpha ij}\mu_{ji}$, since both are obtained by dividing by the same reduced mass parameter $\mu_{ji}$. It is the uncertainty on this $\mu_{ji}$ which dominates the total experimental uncertainty on the modified isotope shift. The improvement obtained through our new value for $^{168}$Yb is clear; the uncertainty is now no longer limited by the mass uncertainty. Furthermore, the 1.7 $\sigma$ deviation on the $^{168}$Yb mass value can clearly be seen. Since the uncertainty on the mass enters the calculation of the modified isotope shifts in similar ways, the effect on the King-plot linearity is, however, minimal: the datapoint shifts approximately parallel to the straight line fit. While a detailed analysis should now be performed, the 3$\sigma$ deviation from linearity will most likely remain unchanged. 

Future, higher-precision measurements of the optical isotope shifts will hopefully serve to better investigate this interesting deviation from linearity. Following the argument presented \cite{Counts2020}, the atomic masses are at present sufficiently precise for King plot studies using isotope shifts with precisions down to $\mathcal{O}$(100 mHz) for $^{168}$Yb $-^{170}$Yb, and $\mathcal{O}$(10 mHz) for the other stable isotope pairs. It should also be noted that this level of precision on the atomic masses can also be obtained for radioactive Yb isotopes with our setup. Thus, the precision of the atomic masses does not restrict future King-plot linearity tests to just the stable isotopes. Finally, once the optical data reaches a $\mathcal{O}$(10 mHz) precision (as was demonstrated recently for e.g. Sr isotopes \cite{manovitz2019precision}), even more precise atomic masses can be obtained with e.g. state-of-the-art cryogenic Penning-trap mass spectrometers \cite{FSU2018,Pentatrap2020}.

\bibliography{mybibfile}

\end{document}